\documentclass[a4paper,fleqn,usenatbib,letters]{mnras}

\usepackage{mathptmx}

\usepackage[T1]{fontenc}
\usepackage{ae,aecompl}

\usepackage{graphicx}	
\usepackage{amsmath}	
\usepackage{amssymb}	

\newcommand{\ha}{H$\alpha$}

\def\lesssim{\mathrel{\hbox{\rlap{\hbox{\lower4pt\hbox{$\sim$}}}\hbox{$<$}}}}
\let\la=\lesssim
\def\gtrsim{\mathrel{\hbox{\rlap{\hbox{\lower4pt\hbox{$\sim$}}}\hbox{$>$}}}}
\let\ga=\gtrsim

\title[AXP 4U 0142+61: optical spectroscopy]{The featureless and non-variable optical spectral energy distribution of AXP 4U 0142+61}

\author[Mu\~noz-Darias, de Ugarte Postigo \& Casares]{T. Mu\~noz-Darias$^{1,2}$, A. de Ugarte Postigo$^{3,4}$ and J. Casares$^{1,2,5}$
\\
$^{1}$ Instituto de Astrof\'isica de Canarias, 38205 La Laguna, Tenerife, Spain\\
$^{2}$ Departamento de astrof\'isica, Univ. de La Laguna, E-38206 La Laguna, Tenerife, Spain\\
$^{3}$ Instituto de Astrof\'{\i}sica de Andaluc\'{\i}a (IAA-CSIC), Glorieta
de la Astronom\'{\i}a s/n, 18008, Granada, Spain.\\
$^{4}$Dark Cosmology Centre, Niels Bohr Institute, Juliane Maries Vej 30, Copenhagen \O, D-2100, Denmark\\
$^{5}$Department of Physics, Astrophysics, University of Oxford, Keble Road, Oxford, OX1 3RH, United Kingdom\\
}

\date{Accepted XXX. Received YYY; in original form ZZZ}

\pubyear{2016}

\begin{document}
\label{firstpage}
\pagerange{\pageref{firstpage}--\pageref{lastpage}}
\maketitle

\begin{abstract}
We present GTC-10.4m spectroscopy and multi-band photometry of the faint ($r\sim 26$) optical counterpart of the anomalous X-ray pulsar 4U 0142+61. The 5000--9000 \AA\ spectrum -- the first obtained for a $magnetar$ -- is featureless, allowing us to set an equivalent width upper limit $EW < 25$~\AA\ to the presence of emission lines in the \ha\ region. Multi-band photometry in the $g$,$r$,$i$,$z$ SDSS bands obtained at different epochs over 12 years shows no significant variability from minutes-to-years time scales. The photometry has been calibrated, for the first time, against the SDSS itself, resulting in solid upper limits to variability ranging from $\sim 0.2$ mag in $i$ (over 12 years) to 0.05 mag in $z$ (over 1.5 years). The shape of the optical + near-infrared (literature values) spectral energy distribution is not well constrained due to the high extinction along the line of sight. Using a Markov Chain Monte Carlo analysis we find that it can be described by a power-law with a spectral index $\beta=-0.7\pm 0.5$ and $E_{(B-V)}=1.5\pm0.4$. We also discuss on the implications of adding hard X-ray flux values from literature to the spectral fitting.  

\end{abstract}




\section{Introduction}
The so-called \textit{magnetars} are isolated neutron stars with dipole magnetic fields in the range $10^{13}$--$10^{15}$ G, as inferred from their spin periods and spin derivatives. Such strong fields are believed to power their X-ray and soft $\gamma$-ray emission \citep{Duncan1992,Thompson1993}, which is also characterized by (X-ray) pulsations and occasional bursts. Magnetar-like activity is associated with two different groups of objects, \textit{Soft Gamma Reapeaters} (SGRs) and \textit{Annomalous X-ray pulsars} (AXPs). SGRs were discovered in 1979 as high-energy transient sources \citep{Mazets1981} which, in some cases, can also be detected as persistent X-ray pulsars. By contrast, AXPs were firstly detected as pulsars in soft X-rays with no optical counterpars \citep{Fahlman1981}. Although some of the differences between these two groups are still not well understood, the common observed properties suggest that AXPs and SGRs are two flavours of the same population of objects. Magnetars are slow rotators with spin periods in the range 2--12 s, but with relatively large period derivatives ($\dot{P} \sim 10^{-13}$ -- $10^{-10}$ s s$^{-1}$ ) pointing to ages between $10^{3}$--$10^{4}$ years (see \citealt*{Mereghetti2015} for a review). 

Alternative models have also been proposed to explain the global properties of magnetars. In particular, the so-called \textit{fossil disc} model advocates for the presence of a cold, accretion disc, possibly generated after the supernova explosion. The combination of accretion and magnetic fields in the upper range of those observed in normal pulsars (a few $10^{13}$ G) could, in principle, explain both the persistent pulsed X-ray emission and the fast spin-down (e.g. \citealt*{Chatterjee2000}).   

AXP 4U 0142+61 (hereafter AXP0142) was discovered by the \textit{Uhuru X-ray observatory} (\citealt{Giacconi1972}; \citealt{Forman1978}). A 8.7s pulsation was found two decades later by \cite{Israel1994}. As generally observed in AXPs, the soft X-rays spectrum (0.5--10 keV) of AXP0142 can be modelled by a strong thermal component with $\kappa T \approx 0.4$ keV, probably associated with the neutron star surface, and a very steep power-law with photon index $\Gamma_\mathrm{soft} \approx$ 3--4 (or, alternatively, an additional black body component). On the other hand, hard X-rays (e.g. 10--200 keV) can be modelled by a single and much flatter power-law with $\Gamma_\mathrm{hard} \approx$ 0.6--1.2 (see \citealt{Rea2007b,Rea2007, denHartog2007,denHartog2008} for X-ray studies).   

The discovery of the faint optical counterpart of AXP0142 by \citet{Hulleman2000}, opened a new window to test the different emission mechanisms in this system and in AXPs in general. The counterpart  has been also detected in the \textit{J, H, K} near infrared bands (nIR; \citealt{Hulleman2004}) and in the mid-infrared (4.5 and 8 $\mu$m; mIR) with the Spitzer Space Telescope (\citealt{Wang2006}). The midIR emission is interpreted as arising from a disc around the central neutron star, which has been suggested to be either truncated and passive \citep{Wang2006} or a gaseous viscously heated accretion disc, thereby contributing to the total radiative power (\citealt{Ertan2007}).   

Important clues can be derived from the evolution of the pulsed fraction ($P_\mathrm{frac}$) as a function of the energy range. This is defined as $P_\mathrm{frac}=(F_\mathrm{max}-F_\mathrm{min}) / (F_\mathrm{max}+F_\mathrm{min}$), where $F_\mathrm{max}$ and $F_\mathrm{min}$ are the maximum and minimum fluxes observed during a pulse period, respectively. In soft X-rays, $P_\mathrm{frac}$ increases from less than $\sim 5\%$ at 0.5--2 keV to at least $\sim 15\%$ at $\sim 8$ keV \citep{Rea2007b}. $P_\mathrm{frac}$ is observed to be in the range $\sim$ 20\%--40\% up to 100 keV \citep{denHartog2008, Tendulkar2015}. Interestingly, a similar $P_\mathrm{frac}$ is found in the optical ($27^{+8}_{-6} \%$ \citealt{Kern2002}; $29\pm 8\%$ \citealt{Dhillon2005}).   

\begin{table}
\label{log_spec}
\caption{Spectroscopy of AXP 4U 0142+61}             
\label{table:spec}      
\centering                          
\begin{tabular}{c c c c}        
\hline\hline                 
Date			& Grism		& Seeing 	&	Exposure time		   \\
\hline
2011-11-28	&	R300R	& 0.9$^{\prime\prime}$	& 2$\times$2400 s\\ 
2011-11-30	&	R300R	& 0.9$^{\prime\prime}$	& 2$\times$2400 s\\ 
2011-12-26	&	R300R	& 0.9$^{\prime\prime}$	& 2$\times$2400 s\\ 
2011-12-27	&	R300R	& 0.9$^{\prime\prime}$	& 3$\times$2400 s\\ 
2011-12-28	&	R300R	& 1.1$^{\prime\prime}$	& 3$\times$2400 s\\ 
2012-01-15	&	R300R	& 0.9$^{\prime\prime}$	& 1$\times$2400 s\\ 
\hline
Combined		&	R300R	&					& 13$\times$2400 s	\\
\hline
\end{tabular}
\end{table}

In this letter, we present what is, to the best of our knowledge, the first optical spectrum of AXP0142, and for extension of a magnetar. We also report an accurate description of the spectral energy distribution (SED) in the optical, including strong constraints to the presence of variability in time scales from minutes to years. Finally, we discuss the broad band SED by considering hard X-ray and nIR data from the literature.

\begin{figure}
\begin{center}
\includegraphics[keepaspectratio,width=9cm]{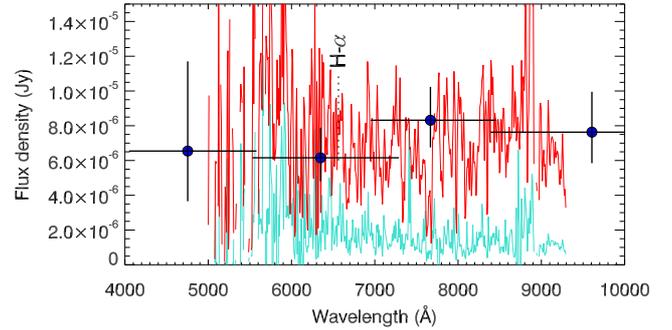}
\caption{Optical spectrum of AXP 4U 0142+61. The spectrum of the source is plotted in red and the spectrum of errors in blue. Photometric values are over-plotted as blue dotes. Fluxes have been corrected for extinction using the mean value from the optical and infrared fit ($E_{(B-V)}=1.5$; see text).}
\label{fig:spectrum}
\end{center}
\end{figure}

\section{Observations and results}
\subsection{Spectroscopy}

Spectroscopy was performed in six epochs between November 2011 and January 2012 using OSIRIS \citep{cepa2000} at the 10.4m GTC telescope (see Tab. \ref{table:spec}). We took 13 exposures of 2400 s for a total of 8.7 h on target, with seeing in the range 0.9--1.1$^{\prime\prime}$. Observations were carried out with the R300R grism, which covers the 4800--10000 {\AA} spectral range with a resolution power of $\sim$350. Following standard procedures -- bias, response, and flux calibration against spectrophotometric standard stars -- we reduced every separate epoch by using a self-made pipeline and IRAF routines. This yielded a series of 2D flux calibrated spectra that were subsequently background corrected, averaged (including a cosmic rejection filter), and stacked to produce the final 1D spectrum (see Fig. \ref{fig:spectrum}). We detect continuum above $\sim$5200 {\AA} and no emission lines. We computed 3$\sigma$ limits to the presence of lines assuming an unresolved line profile ($<850$ km s$^{-1}$). As an example, we obtain an \ha\ flux limit of $4\times10^{-19}$ erg~s$^{-1}$ cm$^{-2}$, corresponding to an equivalent width of 25 {\AA}. For a line breadth of 1500 km s$^{-1}$ the limit would be $8\times10^{-19}$ erg~s$^{-1}$ cm$^{-2}$ or an equivalent width of 47 {\AA}.

\subsection{Photometry}

\begin{figure}
\begin{center}
\includegraphics[width=8cm]{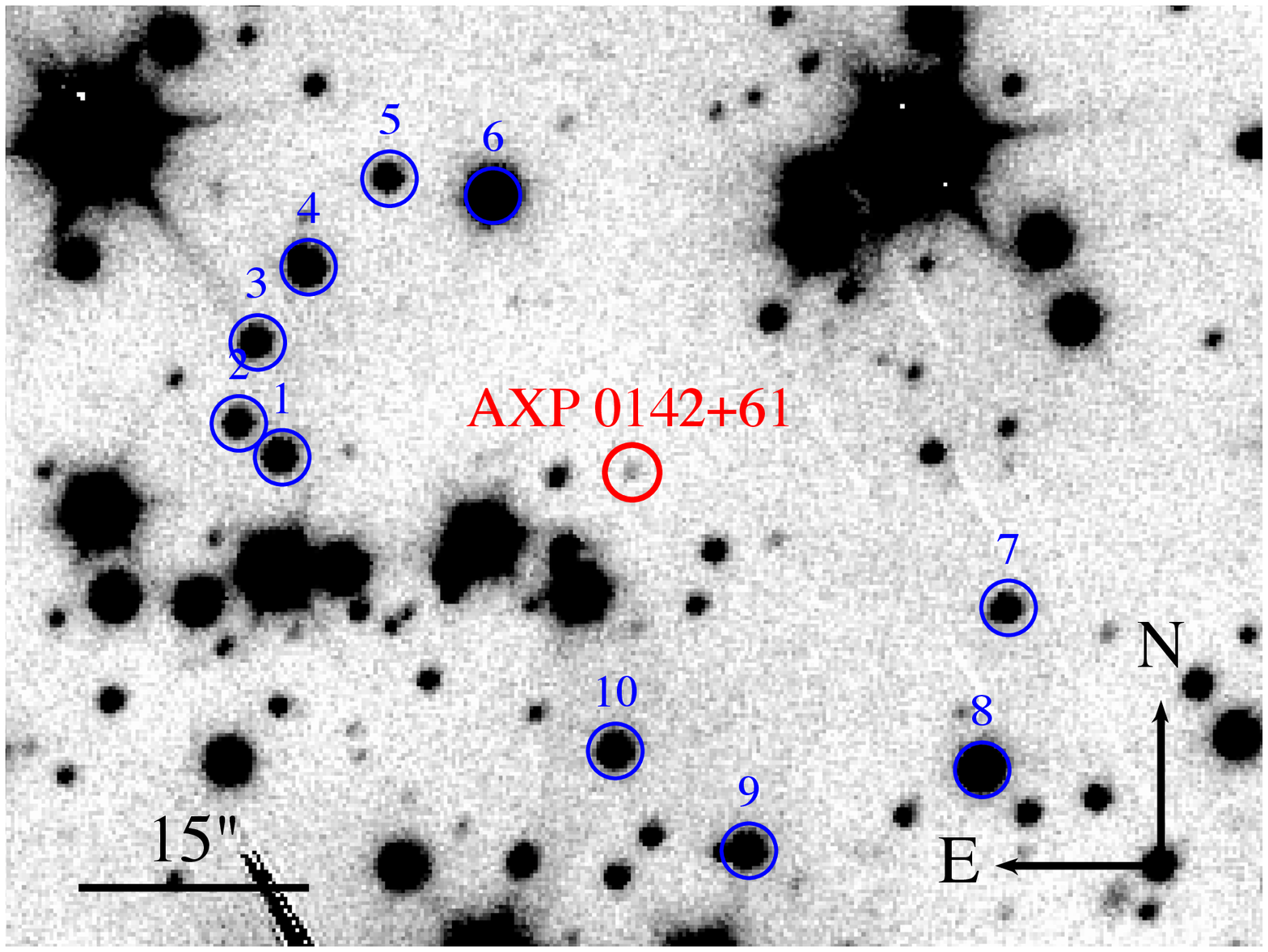}
\caption{$i$-band image of AXP obtained from GTC in August 2013. The field of view is $80^{\prime\prime}\times60^{\prime\prime}$.}
\label{fig:fc}
\end{center}
\end{figure}

GTC+OSIRIS photometry in the $g$, $r$, $i$, $z$ Sloan Digital Sky Survey (SDSS) bands was performed in 9 different epochs spread over 3 years (see Tab. \ref{table:phot}). For absolute flux calibration we used the SDSS reference stars marked in Fig.~\ref{fig:fc} and the magnitudes indicated in Tab.~\ref{table:stars}. We also include the 2002 observations presented in \citet{Dhillon2005}, which were re-calibrated using the same standard stars.

Our $i$-band photometry is stable within 0.2 mag and, therefore, no significant variability is observed over a 11 year baseline between 2002 and 2013. The $g$-band observations, obtained within the same time interval, are also consistent, even though they have larger errors. Observations in the $z$-band within a time span of 1.5 years (January 2012 to August 2013) give consistent results with very little dispersion ($< 0.05$ mag). We note that this is the first study using SDSS bands calibrated against the SDSS survey. The absence of significant variability (e.g. $< 0.05$ mag in $z$) is at odds with previous claims \citep[e.g.][see section \ref{discussion}]{Hulleman2004,Durant2006}.\\
It is worth noting that the observation of 12th January 2012 was obtained 8.51 hr after a $\gamma$-ray flare, indicating that the high-energy activity has a negligible impact on the long-term optical emission. 

\begin{table}
\caption{Photometry of AXP 4U 0142+61. The data were obtained with the OSIRIS camera on the 10.4m GTC, with the exception of the 2002 observations (marked with a *), which were obtained by \citet{Dhillon2005} but re-calibrated for a consistent photometry. Where significant, the zero point errors are indicated in brackets.}             
\label{table:phot}      

\begin{center}                       
\begin{tabular}{c c c c}        
\hline\hline                 
Date			& Filter	& Exposure & Magnitude \\
\hline
2002-09-12*	&	$g$	& 			&	27.4$\pm$ 0.5 \\
2002-09-12*	&	$i$	& 			&	24.59$\pm$ 0.19 \\
2010-07-16	&	$i$	& 120 s			&	24.67$\pm$ 0.24 \\
2011-11-28	&	$i$	& 90 s			&	24.49$\pm$ 0.33 \\
2011-12-26	&	$i$	& 180 s			&	24.58$\pm$ 0.25 \\
2012-01-12	&	$z$	& 6$\times$90 s	&	23.75$\pm$ 0.05(0.23) \\ 
2012-01-15	&	$i$	& 60 s			&	24.48$\pm$ 0.33 \\
2013-08-09	&	$g$	& 12$\times$200 s	&	27.37$\pm$0.25(0.58)  \\
2013-08-09	&	$r$	& 10$\times$200 s	&	25.79$\pm$0.07(0.26)  \\
2013-08-09	&	$i$	& 10$\times$150 s	&	24.55$\pm$0.05(0.22)  \\
2013-08-09	&	$z$	& 10$\times$90 s	&	23.76$\pm$0.07(0.28)  \\
\hline
\end{tabular}
\end{center}
\end{table}

\subsection{The spectral energy distribution}
\begin{figure}
\begin{center}
\includegraphics[keepaspectratio,width=9cm]{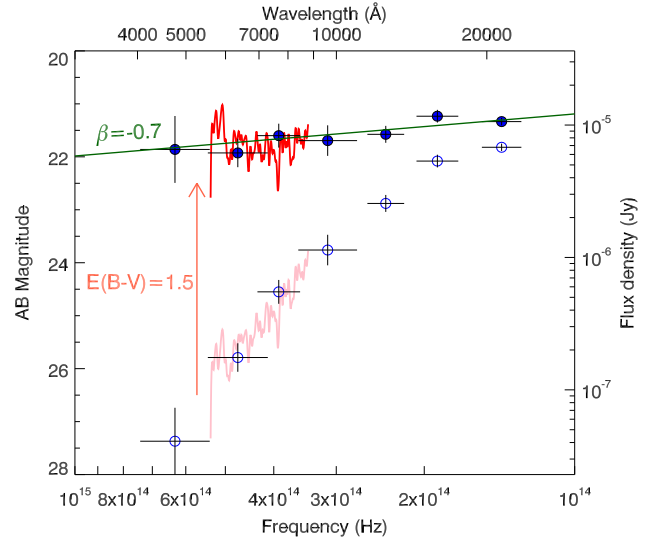}
\caption{Power-law fit ($F_\nu \sim \nu^{\beta}$; green solid line) to the optical and infrared data (dots). Empty circles correspond to (observed) fluxes prior to correcting from extinction ($E_{(B-V)}=1.5$). The spectrum from Fig. \ref{fig:spectrum} is plotted as a red solid line. Note that both the spectral index ($\beta$) and $E_{(B-V)}$ have large associated errors (see text). In the plot we have used the mean values for clarity.}
\label{fig:sedfit}
\end{center}
\end{figure}

In this section we aim at fitting the SED of AXP0142 using simple empirical models. One of the main problems to address during the fitting process is the high interstellar extinction along the line of sight (Milky Way extinction is expected to be $\sim 3.5$ mag in the \textit{V} band). In particular, when fitting the simplest and most standard model, a power-law in the frequency domain ($F_\nu \sim \nu^{\beta}$), the value of the extinction is highly degenerated with the spectral index $\beta$. In order to properly estimate the uncertainties associated with both the extinction $E_{(B-V)}$ and $\beta$, we used the code \textsc{emcee} \citep{Foreman-Mackey2013}, which fits models to data by using a Markov Chain Monte Carlo (MCMC) analysis. We maximized the likelihood function corresponding to a $\chi^2$ distribution, also including the corresponding band widths. In a first step, we used the set of $g$,$r$,$i$,$z$ magnitudes (considering both photometry and zero point errors) taken on 2013-08-09 together with nIR fluxes J=$21.97\pm0.16$, H=$20.66\pm0.12$, $K_s=19.96\pm0.07$ taken on 2004-11-02 with Gemini+NIRI \citep[][see section \ref{discussion}]{Hulleman2004}. We converted these values to the AB system and performed the fit to $F_\nu = K_{n}\nu^{\beta}$ in the log-space, where $K_{n}$ is a normalization variable (see Fig. \ref{fig:sedfit}). The fit also includes an extinction term, $A_{\lambda} = R_{\lambda}*E_{(B-V)}$ where $E_{(B-V)}$ is left variable and the  values $R_{\lambda}$ are fixed to those empirically determined by \cite{Yuan2013} for the SDSS and nIR bands. 
We obtain $\beta=-0.7\pm 0.5$ and $E_{(B-V)}=1.5\pm0.4$. We quote errors using the 16\% and 84\% percentiles of the MCMC equilibrium distribution. Results are correlated in such a way that lower $\beta$ values (i.e. more negative) require lower $E_{(B-V)}$. By applying the standard law $R_V=3.1$ we obtain $A_\mathrm{V}$ in the range 3.1 -- 5.9. On the other hand, the obtained $\beta$ is equivalent to a photon index ($\Gamma=-\beta +1$) in the range 1.2--2.2 . 
We find that these solutions adequately reproduce the data, our best estimation corresponding to $\chi^2_\mathrm{red}$=0.6.\par
 	
\begin{table}
\caption{Photometric reference stars in the field of AXP 4U 0142+61, obtained from the Sloan Digital Sky Survey (DR12). The coordinates of the object are (J2000.0): 01:46:22.44, +61:45:03.3.}             
\label{table:stars}      
\centering                          
\begin{tabular}{c c c c c}        
\hline\hline                 
Star			& $g$-band	& $r$-band	& $i$-band	& $z$-band	\\
\hline
1			& ---				& 21.89$\pm$0.11	& 20.66$\pm$0.05	& 20.00$\pm$0.09	\\
2			& ---				& 22.69$\pm$0.20	& 21.46$\pm$0.10	& 20.83$\pm$0.17	\\
3			& ---				& 22.37$\pm$0.17	& 21.90$\pm$0.15	& 20.25$\pm$0.12	\\
4			& 23.68$\pm$0.30	& 21.23$\pm$0.06	& 20.18$\pm$0.03	& 19.63$\pm$0.06	\\
5			& ---				& 21.98$\pm$0.15	& 20.85$\pm$0.08	& 20.03$\pm$0.12	\\
6			& 20.98$\pm$0.03	& 19.32$\pm$0.01	& 18.45$\pm$0.01	& 17.88$\pm$0.02	\\
7			& ---				& 22.79$\pm$0.26	& 21.06$\pm$0.08	& 20.32$\pm$0.13	\\
8			& 21.94$\pm$0.07	& 20.14$\pm$0.02	& 19.17$\pm$0.02	& 18.45$\pm$0.03	\\
9			& ---				& 22.00$\pm$0.11	& 20.54$\pm$0.05	& 19.68$\pm$0.07	\\
10			& 23.29$\pm$0.27	& 21.34$\pm$0.08	& 20.29$\pm$0.04	& 19.57$\pm$0.07	\\
\hline
\end{tabular}
\end{table}

In a second step we expanded the SED by adding the hard power-law X-ray flux inferred from the \textit{INTEGRAL} ($2.9\pm0.1 \times 10^{-7}$ Jy in 20--150 keV; \citealt{denHartog2008}), \textit{Suzaku} ($3.0\pm0.3 \times 10^{-7}$ Jy in 15--60 keV; \citealt{Enoto2011}) and \textit{NuStar} ($6.3\pm1.1 \times 10^{-7}$ Jy in 0.5--79 keV; \citealt{Tendulkar2015}; within this band we have included only the flux associated with the hard power-law component) X-ray telescopes. This was done to test a possible scenario where the three energy bands (hard X-rays, optical and nIR), which are all highly pulsed (at least the former two), would arise from the same spectral component. Soft X-rays and mid-Infrared data were excluded from the initial fit. In the former case, the spectral analysis together with the low pulsed fraction point towards an extra contribution from the NS surface. Similarly, the excess found by \citet{Wang2006} in the mid-Infrared has been proposed to be associated with an extra disc component.  Applying the same fitting technique we now obtain $\beta=-0.33\pm 0.01$ ($\Gamma \sim $1.3) and $E_{(B-V)}=1.74 \pm0.10$ ($A_\mathrm{V} \sim 5.1$ -- 5.7). We note that given the large separation in frequency between the OIR and the hard X-rays, $\beta$ is fully consistent and similarly well constrained if we exclude any of the hard X-ray fluxes from the fit. For instance, by only using INTEGRAL and SUZAKU we obtain $\beta=-0.33\pm 0.01$ and $E_{(B-V)}=1.76 \pm0.10$.

In a third step we also added the mIR fluxes reported by \citealt{Wang2006} to the SED. We did not find a good power-law fit for any of the above two cases (i.e. considering/excluding the hard X-ray flux). This is consistent with the presence of an extra component in the mIR. Fixing a black body component to a given $T_{BB}$ temperature and correcting the OIR fluxes from its contribution we only find good fits for $T_{BB}\lesssim 700 K$. This $T_{BB}$ is low enough to avoid a strong contribution of the black body to the nIR flux.

\section{discussion}
We have presented the first optical spectrum ever taken of a magnetar. AXP0142, with $r\sim26$, is very challenging even for a 10m class telescope, but we are still able to detect continuum signal throughout the spectral range 5200--10000 {\AA}. The spectrum is featureless and, as an example, we set an upper limit to any possible emission line in the \ha\ region of $EW > 25$ \AA. The spectrum is well matched by the optical photometry, which we find to be non-variable across time-scales from minutes to years. Our variability upper limits range from 0.2--0.3 mag in the bluer and more extinguished bands to $\sim$ 0.05 mag in $z$. The two $z$ time series taken 18 months apart (see Tab. \ref{table:phot})  strongly advocate for a non-variable optical counterpart besides the pulsing activity. This contradicts previous variability claims by \cite{Durant2006}, who collected data presented by different groups and taken with different instruments. Our measurements have been obtained with the same telescope (GTC-10.4m) and instrument (OSIRIS), were taken in the SDSS bands, and are (for the first time) calibrated against the SDSS itself. These results show the importance of performing self-consistent measurements, especially when dealing with objects that are (very) faint and have strong colours such as AXP0142. 

\cite{Hulleman2004} reported the detection of nIR variability. We are not able to test this scenario but we note that calibration might also be an issue here. Indeed, if we only consider the four $K_s$ measurements taken by the same facility (Gemini+NIRI) across 2 years, fluxes are consistent within $\sim$ 0.1 mag (see  table 1 in \citealt{Durant2006}). Also, our most constraining measurement come from the nearby $z$ band at $\lambda_{eff}=9134$~\AA. Nevertheless, two Subaru+IRCS  measurements in $K^\prime$ band taken on consecutive nights are reported to differ by $0.60\pm0.11$ mag \citep{Durant2006}, and nIR variability has also been suggested for other quiescent magnetars \citep{tam2004, testa2008}. Overall, given that  optical variability and emission lines are typical signatures of accretion onto compact objects, our results disfavour any significant contribution from a tentative accretion disc to the optical emission.                        
We find that the OIR SED can be reproduced by a simple power-law model $F_\nu \varpropto\nu^{\beta}$, with  $\beta=-0.7\pm 0.5$. Our self-consistent data do not require of any spectral break between the $B$ ($\lambda_{eff}=4420$~\AA) and $V$ ($\lambda_{eff} = 5400$~\AA) bands, as suggested by \cite{Hulleman2004}. Indeed, our $g$ band ($\lambda_{eff}=4723$~\AA) magnitude is well matched by the power-law fit. Extending the fit by adding hard X-ray measurements results in $\beta=-0.33\pm 0.01$ ($\Gamma \sim$1.3). In both cases, the extinction values are consistent with previous determinations ($2.6\lesssim A_V \lesssim 5.1$; \citealt{Hulleman2004}). The hard X-ray emission, as observed by \textit{INTEGRAL}, \textit{Suzaku} and \textit{NuStar}, is well modelled by a power-law component with $\Gamma_\mathrm{hard}=0.93\pm0.06, 0.89\pm0.11, 0.75\pm0.05$, respectively. These constraints are a few standard deviations away from $\Gamma$=1.3 obtained when jointly fitting nIR, optical and hard X-ray fluxes.  Nevertheless, we note that the $\Gamma_\mathrm{hard}$ values derived by the different X-ray observatories are only marginally consistent between them, and even \textit{INTEGRAL} observations taken at different epochs provide $\Gamma_\mathrm{hard}$ values in the range $0.79\pm0.10$ -- $1.21 \pm 0.16$ (although there is not statistical evidence of variability; \citealt{denHartog2008}).  The above suggests that a spectral model more complex than a simple power-law is probably required to properly describe the broad band SED of AXP0142.

Theoretical studies \citep{Beloborodov2007} propose curvature emission as the origin of the optical and nIR emission from magnetars. In this scenario, a spectral break between the high energies and this spectral regime would be expected. If this is the case, our results suggest that this break occurs at wavelengths shorter than $\sim 5000$ \AA. More observations (e.g. nIR monitoring and timing) and detailed analysis (e.g. full broad band fitting) are necessary to further constrain the origin of the low-frequency emission from magnetars.

\label{discussion}

\section*{Acknowledgements}

We are thankful to the anonymous referee for useful comments. TMD and JC acknowledge support by the Spanish Ministerio de Economia y competitividad (MINECO) under grant AYA2013-42627. JC also ackowledges support by MINECO under grant PR2015-00397. AdUP acknowledges support from a Ram\'on y Cajal fellowship and from the Spanish Ministerio de Economia y competitividad (MINECO) under grant AYA2014-58381. We are thankful to the GTC team that carried out the observations in service mode and to Vik Dhillon for providing the ULTRACAM data presented in Dhillon et al.




\bibliographystyle{mnras}
\bibliography{/Users/tmd/Dropbox/Libreria} 

\begin{thebibliography}{}
\makeatletter
\relax
\def\mn@urlcharsother{\let\do\@makeother \do\$\do\&\do\#\do\^\do\_\do\%\do\~}
\def\mn@doi{\begingroup\mn@urlcharsother \@ifnextchar [ {\mn@doi@}
  {\mn@doi@[]}}
\def\mn@doi@[#1]#2{\def\@tempa{#1}\ifx\@tempa\@empty \href
  {http://dx.doi.org/#2} {doi:#2}\else \href {http://dx.doi.org/#2} {#1}\fi
  \endgroup}
\def\mn@eprint#1#2{\mn@eprint@#1:#2::\@nil}
\def\mn@eprint@arXiv#1{\href {http://arxiv.org/abs/#1} {{\tt arXiv:#1}}}
\def\mn@eprint@dblp#1{\href {http://dblp.uni-trier.de/rec/bibtex/#1.xml}
  {dblp:#1}}
\def\mn@eprint@#1:#2:#3:#4\@nil{\def\@tempa {#1}\def\@tempb {#2}\def\@tempc
  {#3}\ifx \@tempc \@empty \let \@tempc \@tempb \let \@tempb \@tempa \fi \ifx
  \@tempb \@empty \def\@tempb {arXiv}\fi \@ifundefined
  {mn@eprint@\@tempb}{\@tempb:\@tempc}{\expandafter \expandafter \csname
  mn@eprint@\@tempb\endcsname \expandafter{\@tempc}}}

\bibitem[\protect\citeauthoryear{{Beloborodov} \& {Thompson}}{{Beloborodov} \&
  {Thompson}}{2007}]{Beloborodov2007}
{Beloborodov} A.~M.,  {Thompson} C.,  2007, \mn@doi [\apj] {10.1086/508917},
  \href {http://ads.nao.ac.jp/abs/2007ApJ...657..967B} {657, 967}

\bibitem[\protect\citeauthoryear{{Cepa} et~al.,}{{Cepa}
  et~al.}{2000}]{cepa2000}
{Cepa} J.,  et~al., 2000, in {Iye} M.,  {Moorwood} A.~F.,  eds,  Society of
  Photo-Optical Instrumentation Engineers (SPIE) Conference Series Vol. 4008,
  Optical and IR Telescope Instrumentation and Detectors. pp 623--631

\bibitem[\protect\citeauthoryear{{Chatterjee}, {Hernquist}  \&
  {Narayan}}{{Chatterjee} et~al.}{2000}]{Chatterjee2000}
{Chatterjee} P.,  {Hernquist} L.,   {Narayan} R.,  2000, \mn@doi [\apj]
  {10.1086/308748}, \href {http://ads.nao.ac.jp/abs/2000ApJ...534..373C} {534,
  373}

\bibitem[\protect\citeauthoryear{{Dhillon}, {Marsh}, {Hulleman}, {van
  Kerkwijk}, {Shearer}, {Littlefair}, {Gavriil}  \& {Kaspi}}{{Dhillon}
  et~al.}{2005}]{Dhillon2005}
{Dhillon} V.~S.,  {Marsh} T.~R.,  {Hulleman} F.,  {van Kerkwijk} M.~H.,
  {Shearer} A.,  {Littlefair} S.~P.,  {Gavriil} F.~P.,   {Kaspi} V.~M.,  2005,
  \mn@doi [\mnras] {10.1111/j.1365-2966.2005.09465.x}, \href
  {http://ads.nao.ac.jp/abs/2005MNRAS.363..609D} {363, 609}

\bibitem[\protect\citeauthoryear{{Duncan} \& {Thompson}}{{Duncan} \&
  {Thompson}}{1992}]{Duncan1992}
{Duncan} R.~C.,  {Thompson} C.,  1992, \mn@doi [\apjl] {10.1086/186413}, \href
  {http://ads.nao.ac.jp/abs/1992ApJ...392L...9D} {392, L9}

\bibitem[\protect\citeauthoryear{{Durant} \& {van Kerkwijk}}{{Durant} \& {van
  Kerkwijk}}{2006}]{Durant2006}
{Durant} M.,  {van Kerkwijk} M.~H.,  2006, \mn@doi [\apj] {10.1086/507605},
  \href {http://ads.nao.ac.jp/abs/2006ApJ...652..576D} {652, 576}

\bibitem[\protect\citeauthoryear{{Enoto}, {Makishima}, {Nakazawa}, {Kokubun},
  {Kawaharada}, {Kotoku}  \& {Shibazaki}}{{Enoto} et~al.}{2011}]{Enoto2011}
{Enoto} T.,  {Makishima} K.,  {Nakazawa} K.,  {Kokubun} M.,  {Kawaharada} M.,
  {Kotoku} J.,   {Shibazaki} N.,  2011, \mn@doi [\pasj]
  {10.1093/pasj/63.2.387}, \href {http://ads.nao.ac.jp/abs/2011PASJ...63..387E}
  {63, 387}

\bibitem[\protect\citeauthoryear{{Ertan}, {Erkut}, {Ek{\c s}i}  \&
  {Alpar}}{{Ertan} et~al.}{2007}]{Ertan2007}
{Ertan} {\"U}.,  {Erkut} M.~H.,  {Ek{\c s}i} K.~Y.,   {Alpar} M.~A.,  2007,
  \mn@doi [\apj] {10.1086/510303}, \href
  {http://ads.nao.ac.jp/abs/2007ApJ...657..441E} {657, 441}

\bibitem[\protect\citeauthoryear{{Fahlman} \& {Gregory}}{{Fahlman} \&
  {Gregory}}{1981}]{Fahlman1981}
{Fahlman} G.~G.,  {Gregory} P.~C.,  1981, \mn@doi [\nat] {10.1038/293202a0},
  \href {http://ads.nao.ac.jp/abs/1981Natur.293..202F} {293, 202}

\bibitem[\protect\citeauthoryear{{Foreman-Mackey}, {Hogg}, {Lang}  \&
  {Goodman}}{{Foreman-Mackey} et~al.}{2013}]{Foreman-Mackey2013}
{Foreman-Mackey} D.,  {Hogg} D.~W.,  {Lang} D.,   {Goodman} J.,  2013, \mn@doi
  [\pasp] {10.1086/670067}, \href
  {http://ads.nao.ac.jp/abs/2013PASP..125..306F} {125, 306}

\bibitem[\protect\citeauthoryear{{Forman}, {Jones}, {Cominsky}, {Julien},
  {Murray}, {Peters}, {Tananbaum}  \& {Giacconi}}{{Forman}
  et~al.}{1978}]{Forman1978}
{Forman} W.,  {Jones} C.,  {Cominsky} L.,  {Julien} P.,  {Murray} S.,  {Peters}
  G.,  {Tananbaum} H.,   {Giacconi} R.,  1978, \mn@doi [\apjs]
  {10.1086/190561}, \href {http://ads.nao.ac.jp/abs/1978ApJS...38..357F} {38,
  357}

\bibitem[\protect\citeauthoryear{{Giacconi}, {Murray}, {Gursky}, {Kellogg},
  {Schreier}  \& {Tananbaum}}{{Giacconi} et~al.}{1972}]{Giacconi1972}
{Giacconi} R.,  {Murray} S.,  {Gursky} H.,  {Kellogg} E.,  {Schreier} E.,
  {Tananbaum} H.,  1972, \mn@doi [\apj] {10.1086/151790}, \href
  {http://ads.nao.ac.jp/abs/1972ApJ...178..281G} {178, 281}

\bibitem[\protect\citeauthoryear{{Hulleman}, {van Kerkwijk}  \&
  {Kulkarni}}{{Hulleman} et~al.}{2000}]{Hulleman2000}
{Hulleman} F.,  {van Kerkwijk} M.~H.,   {Kulkarni} S.~R.,  2000, \nat, \href
  {http://adsabs.harvard.edu/abs/2000Natur.408..689H} {408, 689}

\bibitem[\protect\citeauthoryear{{Hulleman}, {van Kerkwijk}  \&
  {Kulkarni}}{{Hulleman} et~al.}{2004}]{Hulleman2004}
{Hulleman} F.,  {van Kerkwijk} M.~H.,   {Kulkarni} S.~R.,  2004, \mn@doi [\aap]
  {10.1051/0004-6361:20031756}, \href
  {http://ads.nao.ac.jp/abs/2004A%26A...416.1037H} {416, 1037}

\bibitem[\protect\citeauthoryear{{Israel}, {Mereghetti}  \& {Stella}}{{Israel}
  et~al.}{1994}]{Israel1994}
{Israel} G.~L.,  {Mereghetti} S.,   {Stella} L.,  1994, \mn@doi [\apjl]
  {10.1086/187539}, \href {http://ads.nao.ac.jp/abs/1994ApJ...433L..25I} {433,
  L25}

\bibitem[\protect\citeauthoryear{{Kern} \& {Martin}}{{Kern} \&
  {Martin}}{2002}]{Kern2002}
{Kern} B.,  {Martin} C.,  2002, \nat, \href
  {http://ads.nao.ac.jp/abs/2002Natur.417..527K} {417, 527}

\bibitem[\protect\citeauthoryear{{Mazets} \& {Golenetskii}}{{Mazets} \&
  {Golenetskii}}{1981}]{Mazets1981}
{Mazets} E.~P.,  {Golenetskii} S.~V.,  1981, \mn@doi [\apss]
  {10.1007/BF00651384}, \href {http://ads.nao.ac.jp/abs/1981Ap%26SS..75...47M}
  {75, 47}

\bibitem[\protect\citeauthoryear{{Mereghetti}, {Pons}  \&
  {Melatos}}{{Mereghetti} et~al.}{2015}]{Mereghetti2015}
{Mereghetti} S.,  {Pons} J.~A.,   {Melatos} A.,  2015, \mn@doi [\ssr]
  {10.1007/s11214-015-0146-y}, \href
  {http://ads.nao.ac.jp/abs/2015SSRv..191..315M} {191, 315}

\bibitem[\protect\citeauthoryear{{Rea} et~al.,}{{Rea} et~al.}{2007a}]{Rea2007b}
{Rea} N.,  et~al., 2007a, \mn@doi [\mnras] {10.1111/j.1365-2966.2007.12257.x},
  \href {http://ads.nao.ac.jp/abs/2007MNRAS.381..293R} {381, 293}

\bibitem[\protect\citeauthoryear{{Rea}, {Turolla}, {Zane}, {Tramacere},
  {Stella}, {Israel}  \& {Campana}}{{Rea} et~al.}{2007b}]{Rea2007}
{Rea} N.,  {Turolla} R.,  {Zane} S.,  {Tramacere} A.,  {Stella} L.,  {Israel}
  G.~L.,   {Campana} R.,  2007b, \mn@doi [\apjl] {10.1086/518434}, \href
  {http://ads.nao.ac.jp/abs/2007ApJ...661L..65R} {661, L65}

\bibitem[\protect\citeauthoryear{{Tam}, {Kaspi}, {van Kerkwijk}  \&
  {Durant}}{{Tam} et~al.}{2004}]{tam2004}
{Tam} C.~R.,  {Kaspi} V.~M.,  {van Kerkwijk} M.~H.,   {Durant} M.,  2004,
  \mn@doi [\apjl] {10.1086/426963}, \href
  {http://adsabs.harvard.edu/abs/2004ApJ...617L..53T} {617, L53}

\bibitem[\protect\citeauthoryear{{Tendulkar} et~al.,}{{Tendulkar}
  et~al.}{2015}]{Tendulkar2015}
{Tendulkar} S.~P.,  et~al., 2015, \mn@doi [\apj] {10.1088/0004-637X/808/1/32},
  \href {http://ads.nao.ac.jp/abs/2015ApJ...808...32T} {808, 32}

\bibitem[\protect\citeauthoryear{{Testa} et~al.,}{{Testa}
  et~al.}{2008}]{testa2008}
{Testa} V.,  et~al., 2008, \mn@doi [\aap] {10.1051/0004-6361:20078692}, \href
  {http://ads.nao.ac.jp/abs/2008A%26A...482..607T} {482, 607}

\bibitem[\protect\citeauthoryear{{Thompson} \& {Duncan}}{{Thompson} \&
  {Duncan}}{1993}]{Thompson1993}
{Thompson} C.,  {Duncan} R.~C.,  1993, \mn@doi [\apj] {10.1086/172580}, \href
  {http://ads.nao.ac.jp/abs/1993ApJ...408..194T} {408, 194}

\bibitem[\protect\citeauthoryear{{Wang}, {Chakrabarty}  \& {Kaplan}}{{Wang}
  et~al.}{2006}]{Wang2006}
{Wang} Z.,  {Chakrabarty} D.,   {Kaplan} D.~L.,  2006, \mn@doi [\nat]
  {10.1038/nature04669}, \href {http://ads.nao.ac.jp/abs/2006Natur.440..772W}
  {440, 772}

\bibitem[\protect\citeauthoryear{{Yuan}, {Liu}  \& {Xiang}}{{Yuan}
  et~al.}{2013}]{Yuan2013}
{Yuan} H.~B.,  {Liu} X.~W.,   {Xiang} M.~S.,  2013, \mn@doi [\mnras]
  {10.1093/mnras/stt039}, \href {http://ads.nao.ac.jp/abs/2013MNRAS.430.2188Y}
  {430, 2188}

\bibitem[\protect\citeauthoryear{{den Hartog}, {Kuiper}, {Hermsen}, {Rea},
  {Durant}, {Stappers}, {Kaspi}  \& {Dib}}{{den Hartog}
  et~al.}{2007}]{denHartog2007}
{den Hartog} P.~R.,  {Kuiper} L.,  {Hermsen} W.,  {Rea} N.,  {Durant} M.,
  {Stappers} B.,  {Kaspi} V.~M.,   {Dib} R.,  2007, \mn@doi [\apss]
  {10.1007/s10509-007-9367-1}, \href
  {http://ads.nao.ac.jp/abs/2007Ap%26SS.308..647D} {308, 647}

\bibitem[\protect\citeauthoryear{{den Hartog}, {Kuiper}, {Hermsen}, {Kaspi},
  {Dib}, {Kn{\"o}dlseder}  \& {Gavriil}}{{den Hartog}
  et~al.}{2008}]{denHartog2008}
{den Hartog} P.~R.,  {Kuiper} L.,  {Hermsen} W.,  {Kaspi} V.~M.,  {Dib} R.,
  {Kn{\"o}dlseder} J.,   {Gavriil} F.~P.,  2008, \mn@doi [\aap]
  {10.1051/0004-6361:200809390}, \href
  {http://ads.nao.ac.jp/abs/2008A%26A...489..245D} {489, 245}

\makeatother
\end{thebibliography}




\bsp	
\label{lastpage}
\end{document}